\documentclass[]{aa}
\usepackage{paralist}
\usepackage{graphicx}            
\usepackage{float}                
\usepackage{placeins}             
\usepackage{adjustbox}            
\usepackage{lscape}               
\usepackage{wrapfig}            
\usepackage{epstopdf}         
\usepackage{subfigure}      
\usepackage{booktabs}             
\usepackage{amsmath}              
\usepackage[T1]{fontenc}          
\usepackage{float}              
\usepackage{gensymb}            
\usepackage{rotating}
\usepackage{lineno}  
\usepackage{siunitx}
\usepackage{dcolumn}
\usepackage{rotating}
\usepackage{makecell}
\usepackage{booktabs} 
\usepackage{amsmath} 
\usepackage[justification=justified, singlelinecheck=false]{caption}
\usepackage{txfonts}
\usepackage{natbib}
\usepackage{graphicx}  
\usepackage{amsmath}  
\usepackage{multirow} 
\usepackage{url}
\usepackage{hyperref}
\usepackage{lscape} 
\usepackage{adjustbox}  
\usepackage{caption}
\usepackage{txfonts}

\DeclareRobustCommand{\VAN}[3]{#2}
\let\VANthebibliography\thebibliography
\def\thebibliography{\DeclareRobustCommand{\VAN}[3]{##3}\VANthebibliography}

\DeclareGraphicsExtensions{.ps}
\begin{document}

\title{Iron line diagnostics of the stellar wind in X1908+075}
\titlerunning{Iron line diagnostics in X1908+075}  
\author{Planelles-Villalva, J.$^{1}$, Torrej\'on, J.M.$^{1}$, Rodes-Roca, J.J.$^{1}$, Sanjurjo-Ferr\'{i}n, G.$^{1}$}
\authorrunning{Planelles-Villalva et al.} %

\institute{
$^{1}$Instituto Universitario de F\'{i}sica Aplicada a las Ciencias y las Tecnolog\'ias, Universidad de Alicante, 03690 Alicante, Spain\\
\email{jessica.planelles@ua.es}
}

\abstract{\textit{Aims.} X1908+075 is a supergiant X-ray binary system (SgXB) formed by an evolved OB star and a neutron star (NS) orbiting it with a 4.4-day period. In this paper, we aim to characterize the stellar wind properties, constrain the system geometry, and investigate the origin and variability of the Fe K$\alpha$ fluorescence line. 

\textit{Methods.} We analyzed three \textit{Chandra} HETGS observations taken at different orbital phases. We modeled the continuum using a bulk motion Comptonization (\texttt{Bmc}) model with a partial-covering absorption component. To identify statistically significant features, we performed a blind line search with Monte Carlo simulations that accounted for the look-elsewhere effect. We used the $N_{\mathrm{H}}$ measurements at different orbital phases to constrain the orbital inclination and mass-loss rate through particle swarm optimization.

\textit{Results.}  We detect Fe K$\alpha$ emission in all observations, which remains statistically significant after correcting for multiple trials via Monte Carlo simulations (global $p < 0.005$). Tentative features include a Compton shoulder in the least absorbed observation and Fe K$\beta$ in one observation, suggesting the presence of dense reprocessing material. However, we do not confirm the presence of highly ionized lines (Fe XXV, Fe XXVI). The Fe K$\alpha$ flux correlates positively with the continuum flux, while its equivalent width (EW) shows an anticorrelation with both $N_{\mathrm{H}}$ and orbital phase, opposite to the canonical ``curve of growth'' observed in many HMXBs. Line broadening corresponds to velocities up to $\sim 3000$ km s$^{-1}$, exceeding both the orbital and terminal wind velocities. Modeling the orbital modulation of $N_{\mathrm{H}}$ yields an inclination $i = 46 \pm 3^\circ$ and a donor mass-loss rate $\dot{M}_W = (9.1 \pm 1.6) \times 10^{-7}  M_\odot  \mathrm{yr}^{-1}$. 

\textit{Conclusions.} X1908+075 is a classical wind-fed SgXB in the direct accretion regime, where the NS accretes material from the donor’s dense wind without magnetic or centrifugal barriers. The observed variability in continuum and line emission reflects the inhomogeneous density and ionization structure of the stellar wind. Future high-sensitivity observations with broader orbital phase coverage will help disentangle the relative contributions of fluorescence, scattering, and ionization to the iron line diagnostics.
}

\date{}
   \maketitle
   \nolinenumbers

\section{Introduction}
In high-mass X-ray binaries (HMXBs), a compact object like a NS or black hole (BH) accretes material from a massive companion star, primarily through stellar wind capture or Roche-lobe overflow, producing intense X-ray emission.

The X-ray source X1908+075 belongs to the class of SgXBs, a subclass of HMXBs in which a compact object accretes material from the dense, radiation-driven wind of an evolved OB supergiant companion \citep{ElMellah2017}. In these systems, the mass loss from the donor is governed by a line-driven stellar wind, a quasi-isotropic outflow with high velocity and density, resulting from the resonant absorption of ultraviolet photons by partially ionized metal ions in the stellar atmosphere \citep{Castor1975, Lucy1970}. 

X1908+075 (also known as 4U 1909+07) was first detected by \cite{Forman1978} using data from the \textit{Uhuru} satellite. It has since been observed in surveys with \textit{OSO7}, \textit{Ariel}, \textit{HEAO-1}, \textit{EXOSAT}, and \textit{INTEGRAL}, among others. \cite{Wen2000} analyzed Rossi X-ray Timing Explorer (\textit{RXTE}) All-Sky Monitor data and identified a 4.4-day periodicity in the X-ray intensity, attributed to the binary’s orbital motion.

Spectral and timing analyses by \citet{Molkov2003} using \textit{INTEGRAL} data revealed an X-ray spectrum with a photon index (2–3.9) intermediate between those of BH binaries and X-ray pulsars. However, the compact object's nature remained uncertain. Later, \citet{Levine2004} identified 605 s pulsations in \textit{RXTE} observations, confirming it as a NS. They derived a mass function of 6.1~$M_{\odot}$, constraining the companion's mass and estimating a stellar wind loss rate of $\sim 1.3 \times 10^{-6}$~$M_{\odot}$ yr$^{-1}$.

Near-infrared observations by \cite{Morel2005} identified X1908+075 as an OB-supergiant X-ray binary, locating it at $\sim$7 kpc with an extinction $A_V$ of 16 magnitudes, consistent with X-ray-derived hydrogen column density estimates. Later, \cite{MartnezNez2015} refined the classification to a B0–B3 supergiant with a mass of $15 \pm 6$~$ M_{\odot}$, a mass-loss rate of $ 2.82\times 10^{-7}$ $M_\odot$ yr$^{-1}$,  an effective temperature of 23,000 K, and a distance of $4.9 \pm 0.5$ kpc.

Timing and spectral analyses by \cite{2011A&A...525A..73F} using \textit{INTEGRAL} and \textit{RXTE} found an erratic $\sim$604 s pulse period with energy-dependent profiles, suggesting wind accretion without a stable disk and a strong magnetic field. However, no cyclotron resonance scattering feature (CRSF) was detected. Follow-up \textit{Suzaku} observations analyzed by \cite{Frst2012} confirmed wind accretion behavior, and phase-resolved spectroscopy linked spectral hardness variations to changes in accretion column conditions. \cite{Jaisawal2013} reported an absorption feature at $\sim$44 keV, interpreted as a cyclotron resonance scattering feature (CRSF), suggesting a NS surface magnetic field of $\sim3.8 \times 10^{12}$ G. However, subsequent \textit{NuSTAR} and \textit{Astrosat} observations \citep{Jaisawal2020, Shtykovsky2022} did not confirm this feature. Instead, \cite{Jaisawal2020} estimated a magnetic field exceeding $10^{13}$ G within the framework of quasi-spherical settling accretion theory, although uncertainties in the stellar wind velocity complicated this measurement. \textit{NuSTAR} phase-resolved spectroscopy by \cite{Shtykovsky2022} found no evidence of a CRSF in the 5–55 keV range, constraining the NS magnetic field to either $<5 \times 10^{11}$ G or $>6.2 \times 10^{12}$ G, and revealing pulse profile changes at low energies indicative of multiple emitting regions.

\cite{Corbet2013} discovered superorbital modulation in X1908+075 using Swift BAT, with a period of $\sim$15.2 days, in addition to the known 4.4-day orbital period. This modulation varied over multi-year timescales \citep{Islam2023}, with changes observed in the amplitudes of both the fundamental peaks and their harmonics. Such behavior is consistent with the presence of multiple corotating interaction regions (CIRs) in the winds of the supergiant companions, although a contribution from tidal oscillations to the wind structuring cannot be definitively excluded.

\cite{Islam2023} analyzed Swift and \textit{NuSTAR} data, detecting strong Fe~K$\alpha$ (6.4 keV) and Fe~K$\beta$ (7.1 keV) lines and a 3 ks X-ray flare. The flare showed spectral hardening without increased absorption, consistent with enhanced accretion of clumpy stellar wind structures at low to intermediate rates.

\cite{Torrejn2010} analyzed the {\it Chandra} HETG spectra of this source, focusing on the Fe~K$\alpha$ line and Fe K edge. They detected a Compton shoulder associated with the Fe~K$\alpha$ fluorescence line, suggesting the presence of dense, neutral material surrounding the X-ray source. The narrow width of the Fe K$\alpha$ line, with a full width at half maximum (FWHM) < 5 m$\AA$, indicated that the reprocessing medium did not undergo rapid rotation. The analysis was restricted to the $5-7.7$ keV region, containing both the Fe~K$\alpha$ line and Fe K edge. 

In this paper, we make use of \textit{Chandra} ACIS-S/HETG observations of X1908+075, covering different orbital phases. We use the full spectral range and focus on the lines in the iron complex. Table \ref{tab:stellar_wind_orbital_parameters} summarizes the relevant stellar, wind, and orbital parameters of the source.

We first describe the observational setup and data reduction procedures. We then present the spectral modeling of the continuum and examine its variability across the different observations. Next, we perform a blind search for spectral features in the Fe K band and assess their statistical significance through Monte Carlo simulations that account for the look-elsewhere effect. We then analyze the variability of Fe K$\alpha$ as a function of orbital phase, column density, and source flux. Finally, we discuss the implications of these results for the structure of the stellar wind, the location of the reprocessing material, and the accretion dynamics in SgXBs.

\begin{figure}[h!]
    \centering
    \includegraphics[width=0.5\textwidth]{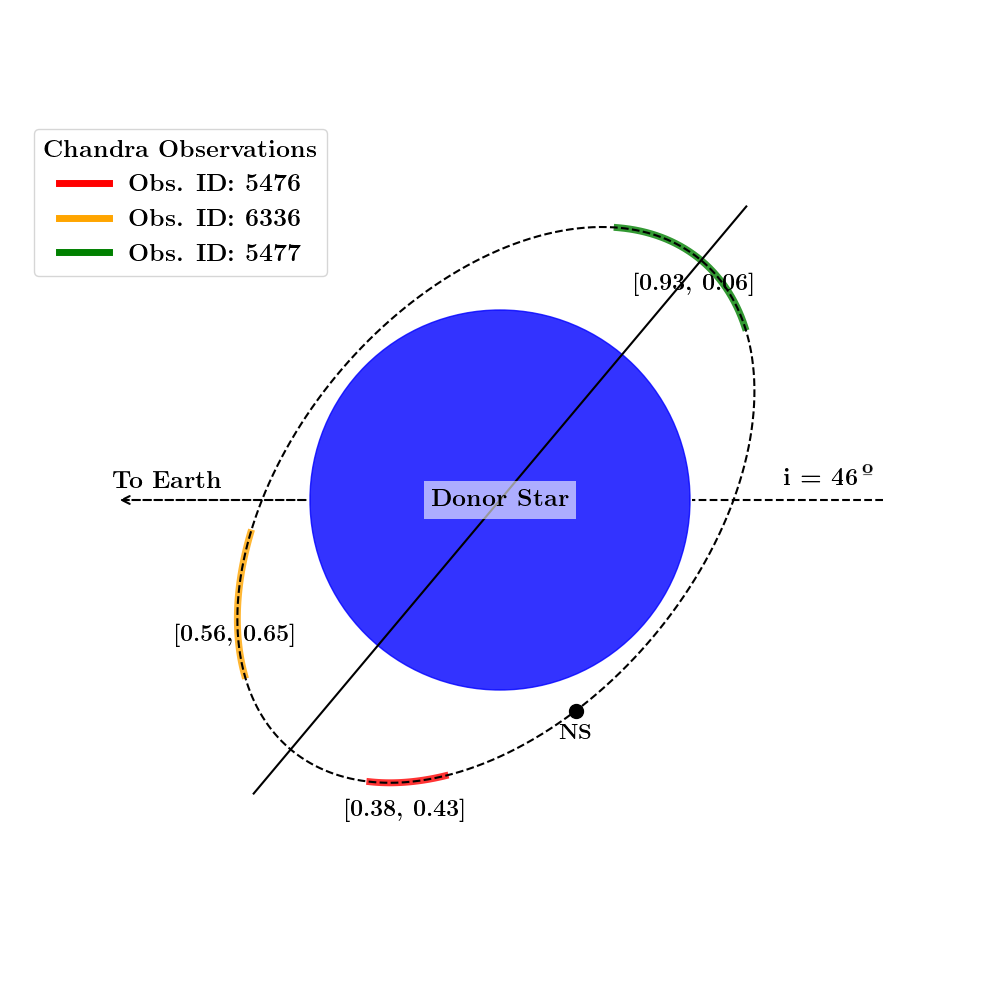}
    \caption{Projected orbit of X1908+075 with marked \textit{Chandra} observations. The diagram shows the donor star (blue) and the binary orbit, including the inclination angle. Observations at different orbital phases are highlighted in distinct colors.} 
    \label{fig:orbit}
\end{figure}

\begin{table}[h!]
\caption{Stellar, wind, and orbital parameters for X1908+075.}
\label{tab:stellar_wind_orbital_parameters}
\centering
\begin{adjustbox}{max width=\columnwidth}
\begin{tabular}{ccc}
\toprule
\toprule
Parameter & Value &  \\ 
\midrule
$a_x$ sin $i$ (lt-s)    & 47.8 ± 0.9          & \citep{Levine2004}      \\ 
$\tau_{90}$ (MJD)     & 52631.383 ± 0.013      & \citep{Levine2004}      \\ 
$P$ (s)              & 604.684 ± 0.001       & \citep{Levine2004}     \\ 
$\dot{P}$ (s s$^{-1}$) & (1.22 ± 0.09) × 10$^{-8}$  & \citep{Levine2004}\\ 
$e$                 & 0.02 ± 0.04    & \citep{Levine2004}          \\ 
$P_{\text{orb}}$ (days)        & 4.4007 ± 0.0009   & \citep{Levine2004}  \\ 
$f(M) (M_{\odot})$ & 6.1 ± 0.4     & \citep{Levine2004}           \\ 
$R_* (R_{\odot})$ & 16       & \citep{Levine2004}      \\ 
$B$ (G) & $<5.6 \times 10^{11}$ or $>6.2 \times 10^{12}$ & \citep{Shtykovsky2022} \\
$v_\text{orb}$ ($\text{km/s}$)& $330 \pm 20  $    & \cite{MartnezNez2015}                     \\ 
$v_\infty $ ($\text{km/s}$)  & $500 \pm 100 $ & \cite{MartnezNez2015}             \\ 
$\beta$ & 1.2 & \cite{MartnezNez2015}\\
$a$ ($\text{cm}$)& $ (1.98 \pm 0.01) \times 10^{12}$  &                   \cite{MartnezNez2015}   \\
$M_* $ ($M_\odot$) & $ 15\pm6$ & \cite{MartnezNez2015}\\
$\dot{M}_W $ ($M_\odot / \text{yr}$) & $ 2.82\times 10^{-7} $ & \cite{MartnezNez2015}\\
$\dot{M}_W $ ($M_\odot / \text{yr}$) & $(9.1 \pm 1.6)\times10^{-7}$ & This work\\
Inclination ($\degree$) & $46 \pm 3$ & This work\\
$v_\text{wind}$  ($\text{km/s}$)  & $190 \pm 40$  &    Derived      \\ 
$R_\text{B}$ ($\text{cm}$)    & $(2.6 \pm 0.4) \times 10^{11}$&     Derived     \\ 
$R_{\text{CO}}$ ($\text{cm}$)& $(1.20 \pm 0.03) \times 10^{10}$  &     Derived                \\ 
$R_\text{A}$ ($\text{cm}$) & $< (3.6 \pm 0.3) \times 10^7 $ or $>(1.43\pm 0.11) \times 10^8  $ &      Derived    \\
$R_\text{M}$ & $< (1.8 \pm 0.13) \times 10^7 $ or $>(7.1\pm 0.5) \times 10^7  $ &      Derived    \\
\bottomrule
\end{tabular}
\end{adjustbox}
\end{table} 
\section{Observations and data analysis}

Our analysis is based on \textit{Chandra} ACIS-S/HETG observations of X1908+075 \citep{Weisskopf2002}. The dataset consists of three observations, obtained at different orbital phases and epochs. A summary of these observations is given in Table \ref{tab:observational_info}. We present a two-dimensional projected plot (Fig. \ref{fig:orbit}) showing the orbital trajectory of the NS around the donor star, based on the orbital parameters listed in Table \ref{tab:stellar_wind_orbital_parameters}. The elliptical orbit is plotted with black dashed lines, taking into account the system’s calculated inclination (see Sect. \ref{sec: inclination}), angle to the periastron and its eccentricity. The donor star is shown in blue at the focus of the ellipse, although the low eccentricity of the system makes this indistinguishable. \textit{Chandra} observations at different orbital phases are highlighted with colored arcs along the orbit, each labeled with its corresponding orbital phase range.

The \textit{Chandra} data were processed and analyzed following standard procedures. Spectra and response files (ARF and RMF) were generated using the \textsc{CIAO} software (v4.15, CalDB 4.15) \citep{Fruscione2006}. To mitigate X-ray loading, we excluded the central region and used an annular extraction region centered on the source. The background was extracted from a similar annulus located outside the source region. First-order dispersions (m = $\pm 1$) from the High Energy Grating (HEG) and Medium Energy Grating (MEG) were extracted and combined. Data in the 0.5–8 keV range were selected for the analysis.

Spectral modeling was carried out using the \textit{Interactive Spectral Interpretation System }\textsc{isis}\footnote{\url{https://space.mit.edu/cxc/isis/}} package \citep{2000ASPC..216..591H}, with additional functionality provided by the \textsc{isisscripts}\footnote{\url{https://www.sternwarte.uni-erlangen.de/isis/}} library. Emission line identification was based on the \textsc{atomdb}\footnote{\url{http://www.atomdb.org/}} and \textsc{nist}\footnote{\url{https://physics.nist.gov/PhysRefData//ASD/lines_form.html}} databases.

\section{Results}

\subsection{Spectral analysis}
\begin{table}[t]
\caption{X1908+075 observational information.}
 \label{tab:observational_info}
\centering
\begin{adjustbox}{max width=\columnwidth}
\begin{tabular}{cccc}
\toprule
\toprule
Obs. ID & \makecell{DATE-OBS \\(\textit{Chandra} Start Date)} & \makecell{Exposure \\(ks)} & \(\phi_{\text{orb}}\) \\ 
\midrule
\textcolor{blue}{\href{https://cda.harvard.edu/chaser/searchOcat.action}{5476}} & 2005-06-27 18:23:08 & 18.73 & [0.38 - 0.43] \\
\textcolor{blue}{\href{https://cda.harvard.edu/chaser/searchOcat.action}{6336}} & 2005-07-11 10:52:53 & 31.42 & [0.56 - 0.65] \\
\textcolor{blue}{\href{https://cda.harvard.edu/chaser/searchOcat.action}{5477}} & 2005-11-26 12:27:37 & 49.00 & [0.93 - 0.06] \\
\bottomrule
\end{tabular}
\end{adjustbox}
\tablefoot{The orbital phase is defined as 0.0 for the epoch that best fits the minimum flux calculated in \cite{Wen2000}:
 \(T_0 = 2\,450\,440.419 \, \text{JD}, \, P = 4.400 \, \text{days}.\)}
\end{table}
We analyzed the full spectrum and examined how the spectral parameters vary with orbital phase. We model the continuum with a \texttt{Bmc} component to describe the NS intrinsic emission. The \texttt{Bmc} model describes the Compton up-scattering of soft seed photons by matter undergoing relativistic bulk motion \citep{1997ApJ...487..834T}. The resulting emission is modified at low energies by an absorption component with a partial covering fraction, \texttt{tbnew-pcf}, which accounts simultaneously for both interstellar and local absorption along the line of sight. 

The \texttt{tbnew-pcf} model implements the Tübingen–Boulder ISM absorption model \citep{Wilms2000}, computing the total photoelectric cross section as the sum of contributions from gas-phase atoms, dust grains, and molecular components. The inclusion of a partial covering fraction allows us to phenomenologically describe the clumpy and inhomogeneous structure of the donor star’s stellar wind. This covering fraction ranges from 0 to 1, where a value of 1 corresponds to full coverage of the X-ray source by the absorber, while lower values represent partial obscuration. In our analysis, this factor turns out to be close to unity, consistent with a NS deeply embedded into the donor's wind.

This model reproduces the observed spectrum well, with a $\chi_{\rm red}^2$ $\sim$1. Below $\sim 2$ keV the model falls below the data. This effect is produced by the soft excess, commonly seen in many HMXBs \citep{2004ApJ...614..881H} and its origin is unclear. One interesting possibility is that it is produced by unresolved emission lines \citep[their Fig. 9]{2025A&A...694A.192S}. 

To explore the soft excess observed between 0.5 and $\leq 2$ keV, we tested the inclusion of an additional thermal component under different absorption configurations. While some configurations marginally improved the fit, they resulted in poorly constrained or physically inconsistent parameters, with no clear trends across observations.

Given the limited number of spectral bins in this energy range and the lack of robust constraints, we do not model this component further in this work. A detailed summary of the tested configurations is provided in Appendix~\ref{appendix:soft_excess}.

\begin{figure*}[h!]
    \centering
    \includegraphics[width=0.99\textwidth]{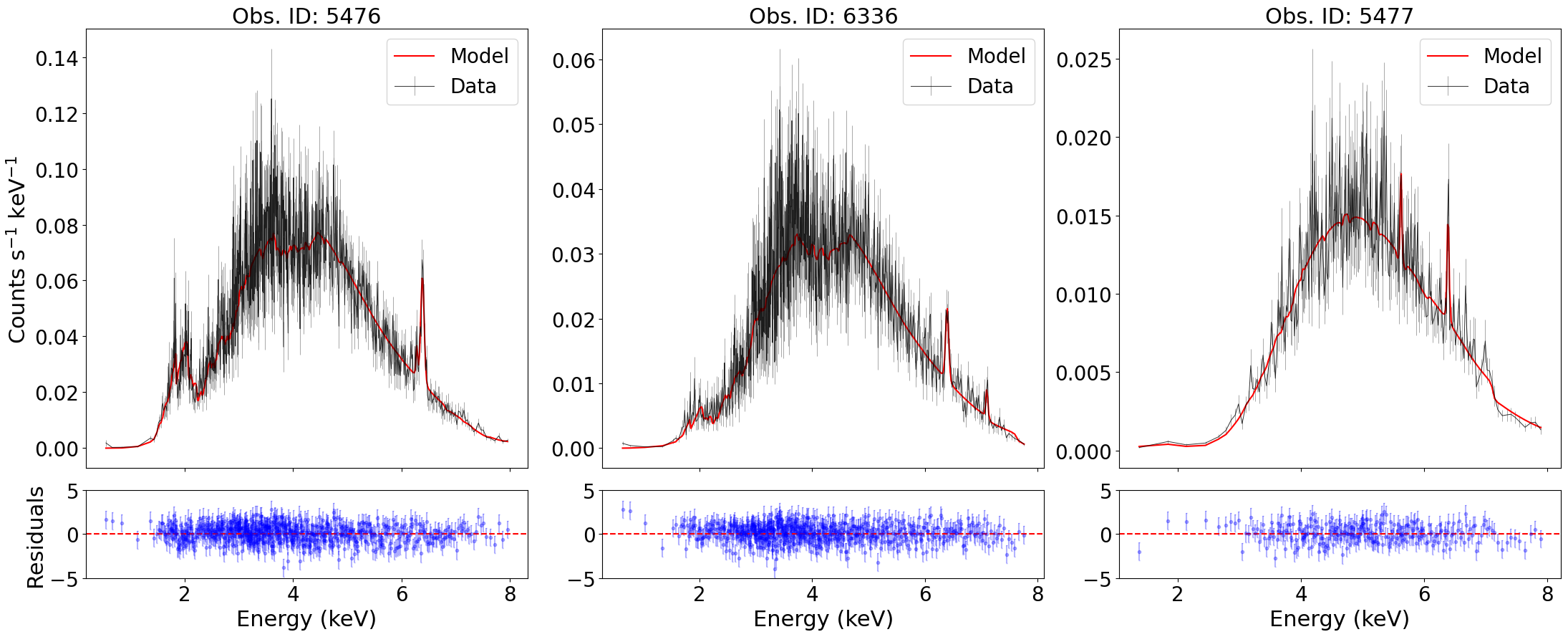}
    \caption{Fitted spectra with the applied absorption and emission model.  
    Residuals are computed as (Data-Model)/$\sigma$, where $\sigma$ is the 1$\sigma$ statistical uncertainty of the observed counts. The residuals therefore represent the deviation of the data from the model in units of standard deviations, indicating how many sigma each spectral bin differs from the model prediction.}
    \label{fig:spectra} 
\end{figure*}

To account for spectral line features, we first performed a blind line search to identify significant emission and absorption features in the K-band. The detected lines were then modeled with Gaussian components (denoted as $\sum G_i$). The complete spectral model is given by:
\begin{align}
F(E) &= \exp(-N_{H}\sigma(E))\times \texttt{pcf}
       \left[\texttt{Bmc}(E) + \sum G_i\right] \nonumber .
\end{align}
Here, $\sigma(E)$ denotes the photoelectric cross section. The \texttt{pcf} parameter ranges from 0 to 1. The best-fit spectral parameters are summarized in Table~\ref{tab:continuum}, and the corresponding spectra together with the best-fit model are shown in Fig.~\ref{fig:spectra}.

It is important to note that our results are consistent with those reported by \citet{torrejon2010}, despite differences in the choice of continuum model and energy range. While \citet{torrejon2010} employed a simple \texttt{Powerlaw} model over a narrower energy range ($5-7.7$ keV), the fitted parameters are in good agreement. In particular, the hydrogen column densities. The $N_{\mathrm{H}}$ values derived by \citet{torrejon2010} ($4.0\pm2.0$, $15\pm4$, and $29\pm13$) are consistent, within uncertainties, with our measurements ($5.5\pm 0.3$, $10.0^{+0.6}_{-0.5}$, and $28.5^{+2.0}_{-1.9}$), in units of $\times10^{22}$ cm$^{-2}$. The largest discrepancy is for Obs. ID 6336. In the restricted range of \citet{torrejon2010}, the $N_{\rm H}$ is obtained mainly from the depth of the K edge while here is driven by the spectral curvature at low energies. In doing so, the depth of the K edge is underestimated (Fig. \ref{fig:spectra}, central panel). It has been not possible to fit both the edge and the curvature well simultaneously. The actual value is probably somewhere in between. However, both $N_{\rm H}$ sets are well correlated with a Pearson correlation coefficient of $0.964$. In any case, the general trends among key parameters, such as the negative trend between $N_{\rm H}$ and the EW of the Fe K$\alpha$ line, are preserved across both studies. Although \citet{torrejon2010} also fitted high-ionization Fe lines, these values were not discussed in their analysis; here, we use our own measurements consistently in the discussion that follows.

\begin{table*}
    \centering
    \caption{Best-fit continuum spectral parameters for X1908+075.}
    
    \begin{adjustbox}{max width=\textwidth}
    \begin{tabular}{l c c c}
    
    \hline
    Parameter & Obs. ID: 5476 & Obs. ID: 6336 & Obs. ID: 5477 \\
    \hline
    $N_{\rm H}$ ($\times 10^{22}$ cm$^{-2}$) & $5.5\pm 0.3$ & $10.0^{+0.6}_{-0.5}$ & $28.5^{+2.0}_{-1.9}$ \\
    pcf & $0.994\pm0.004$ & $0.9947\pm0.0019$ & $0.9935^{+0.0014}_{-0.0016}$\\
    $A_N$ (\texttt{Bmc}) & $\left(4.27^{+0.21}_{-0.19}\right)\times10^{-3}$ & $\left(1.95^{+0.09}_{-0.08}\right)\times10^{-3}$ & $\left(2.5^{+0.4}_{-0.3}\right)\times10^{-3}$\\
    kT (keV) (\texttt{Bmc}) & $2.01^{+0.08}_{-0.07}$ & $1.85\pm0.08$ & $2.5^{+0.3}_{-0.2}$ \\
    $\alpha$ (\texttt{Bmc}) & \multicolumn{3}{c}{4 (fixed)} \\
    $F_X$ ($\times10^{-10}$ erg s$^{-1}$ cm$^{-2}$) & $2.70^{+0.13}_{-0.12}$ & $1.30^{+0.06}_{-0.05}$ & $1.19^{+0.17}_{-0.12}$ \\
    $L_X$ ($\times 10^{35}$ erg s$^{-1}$) & $7.8\pm 1.6$ & $3.7\pm 0.8$ & $3.4^{+0.9}_{-0.8}$ \\
    $\chi^2_{\rm red}$ (dof) & 1.01 (665) & 1.07 (612) & 1.07 (258) \\
    
    \bottomrule
    \end{tabular}
    \end{adjustbox}

    \tablefoot{We report the 0.5–8 keV flux and derived luminosity $L_X = 4\pi d^2 F_X$ using $d = 4.9 \pm 0.5$ kpc \cite{MartnezNez2015}. Uncertainties are at 90\% confidence.}
    
    \label{tab:continuum}
\end{table*}

\subsection{Blind line search}
To search for narrow spectral features, we performed a blind line scan in wavelength space using a grid-search technique (e.g., \citealt{Pinto2017, Kosec_2018}). Starting from the best-fit continuum model, we added a narrow Gaussian component with fixed centroid and width, leaving the normalization free to account for both emission and absorption features. The centroid was stepped across the 1.55–2.48 $\AA$ range (5–8 keV) in increments of 0.002 $\AA$, comparable to the instrumental resolution. For each trial, we recorded the improvement in the fit statistic, $\Delta \chi^2$, and measured the EW, identifying candidate features from local $\Delta \chi^2$ improvements.

To account for the look-elsewhere effect inherent to blind searches, we performed Monte Carlo simulations by generating 1000 spectra based on the best-fit continuum model and repeating the full grid-search procedure for each simulated dataset, recording the maximum $\Delta \chi^2$ value. The global significance of each feature was then estimated by comparing the observed $\Delta \chi^2$ with the distribution of simulated maxima. Adopting a threshold of $p < 0.005$, only the Fe~K$\alpha$ emission line is statistically significant, while all other candidate features are considered tentative after accounting for the look-elsewhere effect.

\begin{figure*}[h!]
    \centering
    \includegraphics[width=0.99\textwidth]{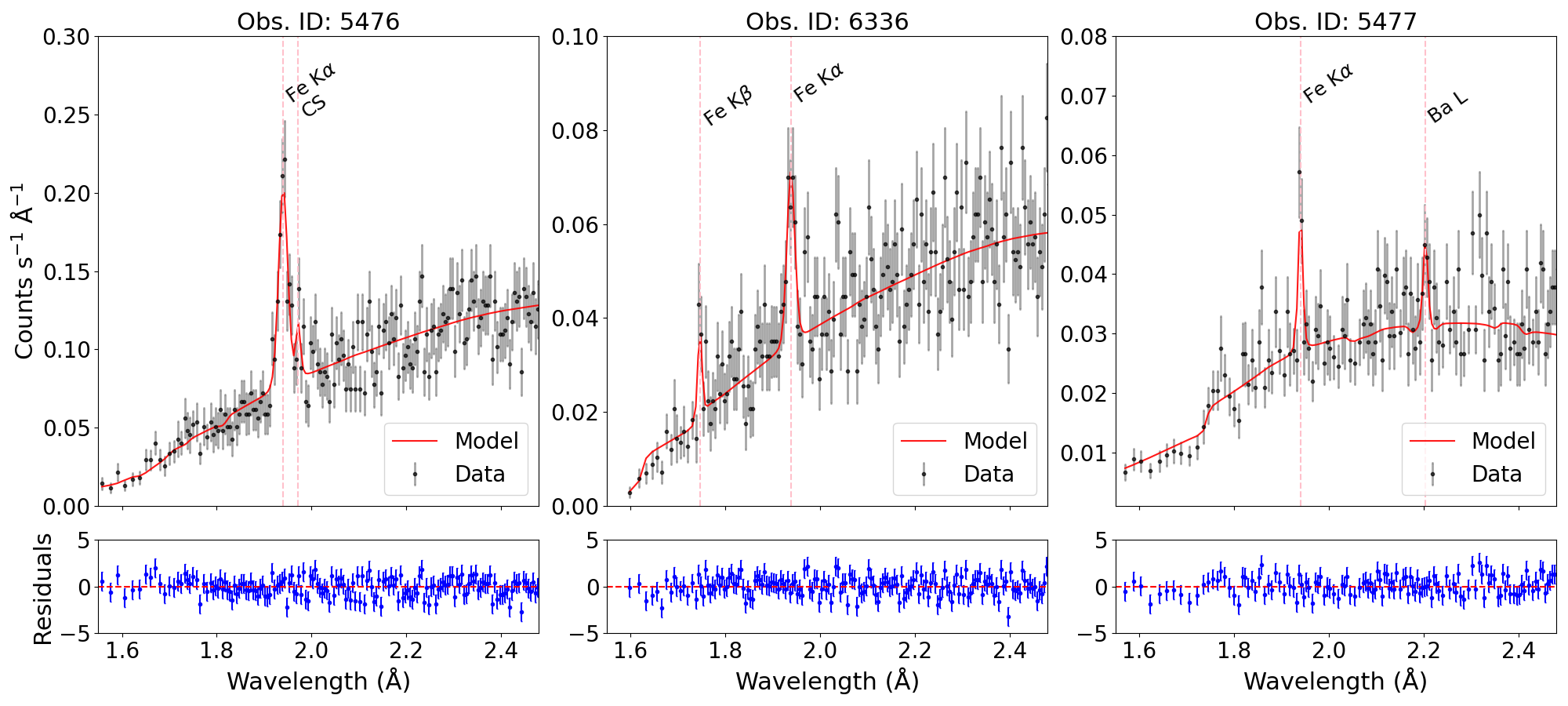}
    \caption{Fitted \texttt{Gaussian} components representing the emission lines detected in the blind line search after accounting for the look-elsewhere effect in X1908+075. We show both statistically significant lines and tentative candidates. The red dashed lines indicate the laboratory centroid wavelength of the labeled transitions.} 
    \label{fig:lines}
\end{figure*}

The spectrum was subsequently modeled including Gaussian components to characterize these features. The Fe~K$\alpha$ line was fitted with free centroid, width, and normalization, while the remaining tentative features were modeled with fixed width. These features are found at wavelengths broadly consistent with Fe~K$\beta$, Ba L\footnote{The centroid wavelength of this feature is consistent with the Ba L line listed in the NIST Atomic Spectra Database (\url{https://physics.nist.gov/asd}). Given its low statistical significance and the fact that such transitions are not commonly observed in HMXBs, we do not attempt a detailed interpretation}, and a possible low-energy structure adjacent to Fe~K$\alpha$ that may be interpreted as a Compton shoulder.
A summary of these features and their statistical properties is presented in Table~\ref{tab:lines}. A detailed description of the blind search procedure, together with the $\Delta \chi^2$ and EW scans and the Monte Carlo significance assessment, is provided in Appendix~\ref{appendix:linesearch}. The corresponding spectral fits including the Gaussian components are shown in Fig.~\ref{fig:lines}, where both statistically significant and tentative lines are displayed. Given that only the Fe~K$\alpha$ line is detected with high statistical significance, the following analysis focuses on its properties and their relation to the physical parameters of the system.

\begin{table*}[h!]
\caption{Blind line search results for X1908+075 in the Fe K band.}
\resizebox{\textwidth}{!}{

\begin{tabular}{cccccccccc}
\hline
\hline
Obs. ID & $\lambda$ ($\AA$) & $E$ (keV) & Line Id.\footnote{Source: \url{https://xdb.lbl.gov/Section1/Table_1-3.pdf}} 
& I (Photons s$^{-1}$ cm$^{-2}$) 
& $\sigma$ (m$\AA$) 
& EW (m$\AA$) 
& $\Delta \chi^2$ 
& $p$-value 
& $\sigma$ \\
\hline

5476 & 1.940 & 6.39 & Fe K$\alpha$  & $\left(5.1\pm0.9\right)\times10^{-4}$ &  $8.1^{+2.4}_{-2.3}$ & $41\pm 7$ & 102.6 & $<0.001$ & $>3.29$\\
5476 & 1.972 & 6.29 & Compton Shoulder & $\left(7\pm5\right)\times10^{-5}$ & -- & $6\pm 4$ & 9.6 & 0.318 & 1.00 \\

\hline

6336 & 1.938 & 6.40 & Fe K$\alpha$  & $\left(1.4\pm0.4\right)\times10^{-4}$ & $7^{+4}_{-3}$ & $23.3^{+6.8}_{-6.5}$ & 33.2 & 0.001 & 3.29 \\
6336 & 1.746 & 7.11 & Fe K$\beta$  & $\left(8\pm5\right)\times10^{-5}$ & -- & $14\pm 7$ & 10.7 & 0.240 & 1.17 \\

\hline

5477 & 1.940 & 6.39 & Fe K$\alpha$ &$\left(6.5\pm2.5\right)\times10^{-5}$ & 2 & $12\pm 4$ & 19.6 & 0.004 & 2.88  \\
5477 & 2.204  & 5.63 & Ba L & $\left(3.3\pm1.8\right)\times10^{-5}$ & -- & $7\pm 4$ & 9.5 & 0.247 & 1.16 \\

\bottomrule
\end{tabular}
}
\label{tab:lines}
\tablefoot{Line centroids are fixed; Fe K$\alpha$ has free width and normalization, while other lines have fixed width (0.002\,\AA). Uncertainties are at 90\% confidence.}
\end{table*}
\begin{table}[h!]
    \centering
    \caption{Best-fit parameters of the Fe~K$\alpha$ line.}
    \begin{adjustbox}{max width=\textwidth}
    \begin{tabular}{ccccc}
    \toprule
    \toprule
    Obs. ID & $\sigma$ (m$\AA$) & FWHM (m$\AA$) & $v_{\mathrm{FWHM}}$ (km s$^{-1}$) \\
    \hline
    5476 & $8.1\pm 2.3$ & $19^{+6}_{-5}$ & $3000\pm 800$ \\
    6336 & $6^{+4}_{-3}$ & $15^{+8}_{-6}$ & $2400^{+1200}_{-1000}$ \\
    5477 & $2.0$ & $4.7$ & $700$ \\
    \bottomrule
    \end{tabular}
    \end{adjustbox}
    \tablefoot{For Obs ID 5477, $\sigma$ was fixed to $2\times10^{-3}$\,\AA\ and no uncertainties are given for this case.}
    \label{tab:line_kalpha}
\end{table}
\subsubsection{Fe~K$\alpha$ analysis}
\begin{figure}[htpb]
    \centering
    \includegraphics[width=0.5\textwidth]{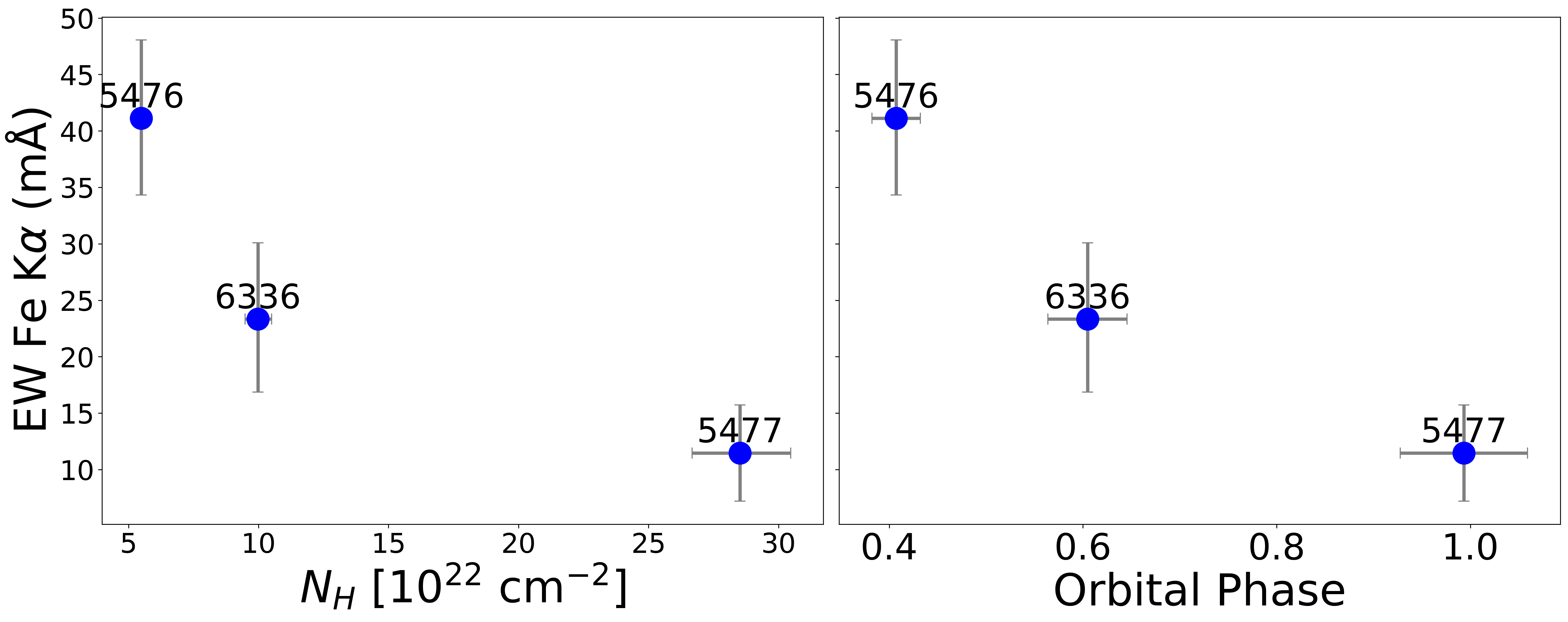}
    \caption{
    Left: EW of the Fe~K$\alpha$ line as a function of the hydrogen column density $N_{\rm H}$.
    Right: EW of the Fe~K$\alpha$ line as a function of the Orbital Phase.
    }
    \label{fig:Fe_kalpha_vs_flux}
    \label{fig:EW_vs_flux}
\end{figure}
\begin{figure}[htpb]
    \centering
    \includegraphics[width=0.5\textwidth]{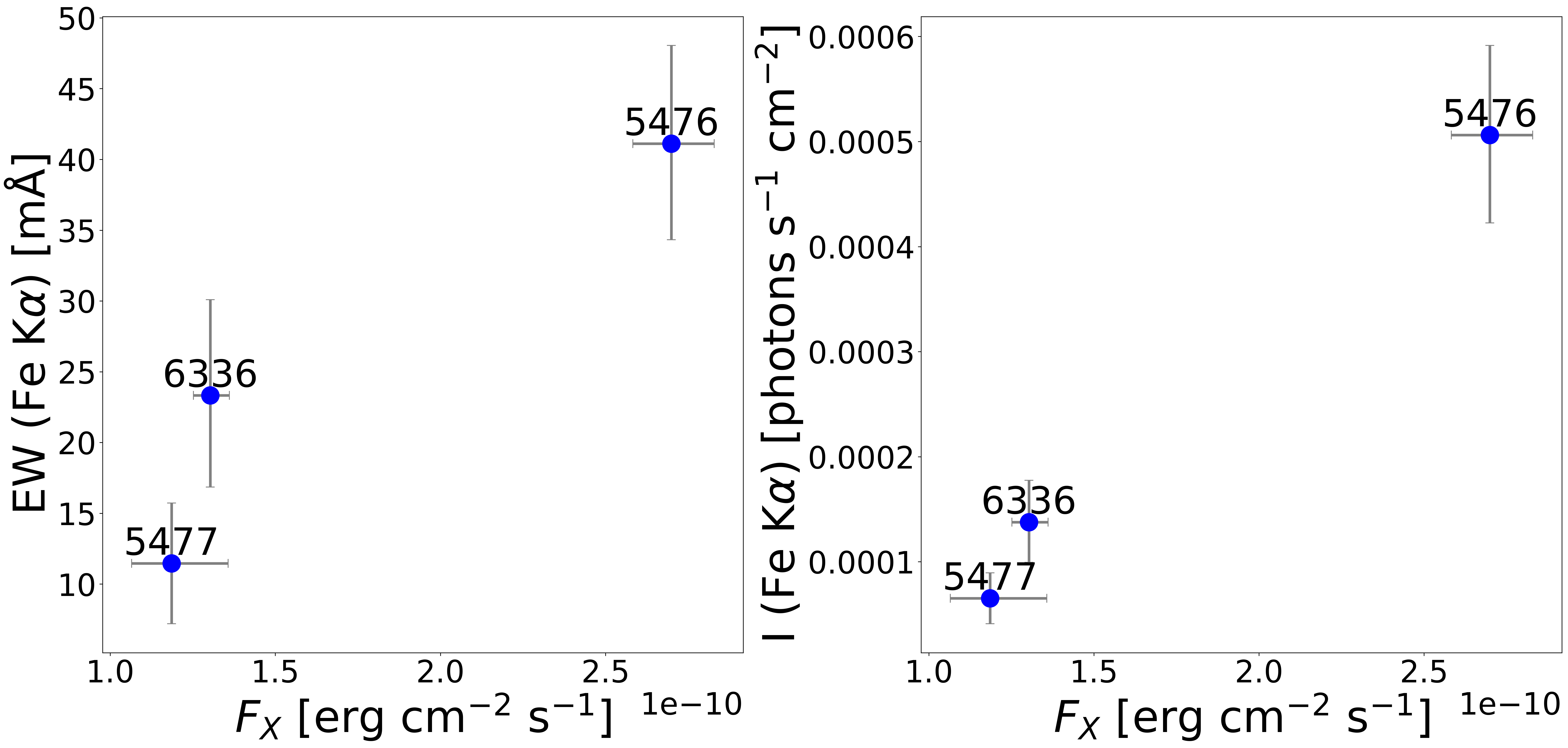}
    \caption{
    Left: EW of the Fe~K$\alpha$ line as a function of the unabsorbed continuum flux, in logarithmic scale.
    Right: Flux of the Fe~K$\alpha$ line as a function of the unabsorbed continuum flux, in logarithmic scale. 
    }
    \label{fig:EW_NH}
    \label{fig:EW_OP}
\end{figure}
The derived line width $\sigma$, FWHM, and velocity broadening ($v_{\mathrm{FWHM}}$) are listed in Table~\ref{tab:line_kalpha}.

Velocity broadening and narrowness of the Fe~K$\alpha$ line. \label{sec:dens}
For a \texttt{Gaussian} profile, the FWHM is related to the standard deviation $\sigma$ by the expression:
\begin{equation}
    \text{FWHM} = 2\sqrt{2\ln 2} \, \sigma \approx 2.355 \, \sigma.
\end{equation}

This defines the characteristic width of the emission line. The corresponding velocity width can be derived from the Doppler relation:
\begin{equation}
    v_{\text{FWHM}} = \frac{\text{FWHM}}{\lambda_0} c,
\end{equation}
where $v_{\text{FWHM}}$ represents the velocity dispersion, $\lambda_0$ is the central wavelength of the emission line (expressed in the same units as the FWHM), and $c$ is the speed of light.

The FWHM of the Fe~K$\alpha$ line is $19^{+6}_{-5}$ m\AA\ in observation 5476 and $15^{+8}_{-6}$ m\AA\ in observation 6336, indicating a relatively broad profile. In contrast, the line appears narrower in observation 5477, with an unresolved FWHM. We thus fix it to 2 m\AA.

The corresponding $v_{\rm FWHM}$ values are $(3000\pm 800)$ km~s$^{-1}$ and $(2400^{+1200}_{-1000})$ km~s$^{-1}$ for observations 5476 and 6336, respectively, about an order of magnitude larger than the stellar wind terminal velocity deduced from the analysis of IR data ($\sim 500$ km~s$^{-1}$). In observation 5477, however, the  $v_{\rm FWHM}=700$ km~s$^{-1}$, which is comparable to $v_{\infty}$.

\paragraph{Fe~K$\alpha$ variability.}
In Fig. \ref{fig:Fe_kalpha_vs_flux} left panel, we present the unabsorbed continuum flux in the 0.5–8 keV range, excluding the flux of the Fe~K$\alpha$ emission as well as the other emission lines, plotted against the Fe~K$\alpha$ flux. The line flux exhibits a positive trend with the continuum flux. This is expected: the surrounding material shows more fluorescence with increasing illumination. 

However, Fig. \ref{fig:EW_NH} reveals a negative trend between the EW of the Fe~K$\alpha$ line and the $N_{\rm H}$. As $N_{\rm H}$ increases, the EW systematically decreases. A similar trend is observed with the EW of the Fe~K$\alpha$ line decreasing with the orbital phase. This behavior is the opposite observed for other HMXBs \citep{torrejon2010, gimenez2015}. Indeed, Fe~K$\alpha$ in HMXBs display a growing EW with $N_{\rm H}$, the so called curve of growth.

The origin and variability of the Fe~K$\alpha$ emission line are examined in detail in Sect. \ref{section:4.1}, including its dependence on the $N_{\rm H}$, orbital phase, and the unabsorbed continuum flux, as well as the implications of line broadening and ionization state of the reprocessing material.

\paragraph{Compton shoulder.} In our analysis, we tentatively identify the CS at the $1\sigma$ significance level in observation 5476, the less absorbed one, where it appears on the red wing of the Fe~K$\alpha$ emission line. The behavior of the CS associated with the Fe~K$\alpha$ line, as well as the absence of this feature in the other observations, is discussed in detail in Sect.~\ref{section:4.1}. 

\subsubsection{Fe~K$\beta$} In observation 6336, the emission line Fe~K$\beta$ is considered tentative at $\sim$ 1$\sigma$. We measured the Fe~K$\alpha$/Fe~K$\beta$ intensity ratio as $0.6\pm 0.4$, which shows deviations from theoretical predictions. These deviations, along with possible explanations—including the influence of the Fe~K absorption edge—are discussed in detail in Sect.~\ref{section:4.1}.
\section{Discussion}
\label{section:4.1}
To characterize the interaction between the NS and the stellar wind, we calculated the characteristic radii that determine the accretion regime of the system. These include the Bondi--Hoyle radius, $R_{\rm B}$, which defines the region where the gravitational potential of the NS dominates over the kinetic energy of the incoming wind \citep{Bondi1944}; the corotation radius, $R_{\rm co}$, at which the NS spin angular velocity equals the local Keplerian frequency \citep{pringle1972accretion}; and the Alfvén radius, $R_{\rm A}$, where the magnetic pressure of the NS balances the ram pressure of the accreting material. The wind velocity at the orbital separation was computed using a standard $\beta$-velocity law \citep{Castor1975}, adopting the stellar and orbital parameters listed in Table~\ref{tab:stellar_wind_orbital_parameters}.

For the Alfvén radius, $R_{\rm A}$, we followed the formulation of \citet{Bozzo2008} for wind-fed accretion. In this framework, $R_{\rm A}$ depends on the NS surface magnetic field $B$, the mass accretion rate $\dot M$, and the NS mass and radius. We adopted the NS parameters from previous studies, while we used the mass accretion rate parameter computed in this work with the \texttt{xraybinaryorbit} package. In our calculations, we derived the Alfvén radius using
\begin{equation}
R_{\rm A} = \left( \frac{\mu^4}{2 \, G \, M_{\rm NS} \, \dot{M}^2} \right)^{1/7}, \quad
\mu = B \, R_{\rm NS}^3,
\end{equation}
which is algebraically equivalent to the expression given by \citet{Zhang2006}. We then estimated the effective magnetospheric radius as $R_{\rm M} = 0.5\, R_{\rm A}$. We adopted the magnetic field values reported by \citet{Shtykovsky2022} as the intrinsic surface field. We did not apply any correction for gravitational redshift (approximately $1+z_g \simeq 1.4$ at the NS surface), which would slightly reduce the effective field measured at infinity and, consequently, the inferred magnetospheric radii.

We summarize in Table~\ref{tab:stellar_wind_orbital_parameters} the resulting characteristic radii. We obtained  $R_{\rm M} < R_{\rm co} < R_{\rm B}$ which places X1908+075 in the direct accretion regime \citep{Fornasini2024}. In this regime \citep{Bozzo2008}, the NS magnetosphere is fully contained within the corotation and Bondi radii, allowing material from the stellar wind to accrete directly onto the NS surface. There is no centrifugal or magnetic barrier to the inflow, so the accretion rate is primarily determined by the local wind density and velocity. Consequently, the system can sustain relatively steady X-ray emission, characteristic of classical wind-fed HMXBs.

\paragraph{System inclination and mass loss rate.}
\label{sec: inclination}
Our analysis reveals a significant modulation of the X-ray absorption $N_{\rm H}$ with orbital phase, reaching its peak near $\phi_{\text{orb}} = 1$. Such a sharp increase in $N_{\rm H}$ has been reported in previous studies. \cite{Levine2004} observed a significant rise (by a factor of 2 or more) in photoabsorption around orbital phase $\phi_{\text{orb}} = 1$, reaching values of $N_{\rm H} \geq 30 \times 10^{22}$ cm$^{-2}$. Similarly, \cite{Walter2006} documented comparable variations in other SgXBs, such as IGR J16318-4848, where $N_{\rm H}$ increased by a factor of $10^2$ between different observations. 

The hydrogen column density, $N_{\mathrm{H}}$ that X-ray emission must traverse before reaching the observer, in a smooth wind approximation, using the CAK models \citep{Castor1975}, is determined by several factors. These include the mass-loss rate, $\dot{M}$, of the donor star and the orbital configuration of the system, characterized by the orbital separation, eccentricity, and argument of periapsis. The value of $N_{\mathrm{H}}$ depends on the orientation of the system relative to the observer, in particular the orbital inclination and orbital phase. Since inclination and eccentricity modulate the line-of-sight path length through the stellar wind, they induce phase-dependent variations in the observed column density. In the limiting case of zero inclination and zero eccentricity, no orbital modulation of $N_{\mathrm{H}}$ would be expected. In the present study, we have three observations obtained at different orbital phases, each yielding a distinct measurement of $N_{\mathrm{H}}$. We use these phase-resolved measurements to constrain both the orbital inclination and the system’s mass-loss rate, $\dot{M}$. For the fit, we adopted the orbital parameters from \cite{MartnezNez2015}. 

We restricted the lower limit for the inclination to the non-eclipsing range $38^{\circ} \le i \le 73^{\circ}$. The upper limit is the one derived by \citet{Levine2004}. The mass‐loss rate was allowed to vary between $10^{-5}$ and $10^{-8}\,M_\odot\,\mathrm{yr}^{-1}$. 

The complexity of the model constrains the use of a classical least-squares approach. \cite{Levine2004} used this method but the eccentricity and angle to the periapsis were not taken into account in their approach, providing a $38^{\circ}-73^{\circ}$ range for the inclination. Instead, we fitted the data using a particle swarm optimization (PSO) algorithm. In PSO, each particle represents a potential solution to an optimization problem. The particles move through the solution space to find the optimal parameter configuration \citep{10.1162/EVCO_r_00180}. The stochastic nature of the method provides slightly different values for each iteration, and the errors given for each parameter are the standard deviation of the values obtained after 10 iterations. The fitting function used is publicly available in the Python package \texttt{xraybinaryorbit} \citep{SanjurjoFerrn2024}.

Our fit predicts column densities of $N_{\mathrm{H}} = 5.5$, $7.9$, and $27.3 \times 10^{22}$~atoms~cm$^{-2}$, which are in very good agreement with the observed fitted values after subtracting the interstellar contribution ($N_{\rm H}$; Table~\ref{tab:continuum}).

The result suggests an inclination of 46 $\pm$ 3$^\circ$ and a mass loss rate of $(9.1 \pm 1.6)\times10^{-7}\,M_\odot\,\mathrm{yr}^{-1}$, in agreement, within a factor of $\sim 3$, with the one given by \cite{MartnezNez2015} based on the analysis of the infrared spectrum. These values are included in Table \ref{tab:stellar_wind_orbital_parameters}. For a detailed description of the model and other predicted parameters, see Appendix \ref{appendix:nh_ps_fit} and Table~\ref{tab:orbital_params_ps}.

\begin{figure*}[h!]
    \centering
    \includegraphics[width=\textwidth]{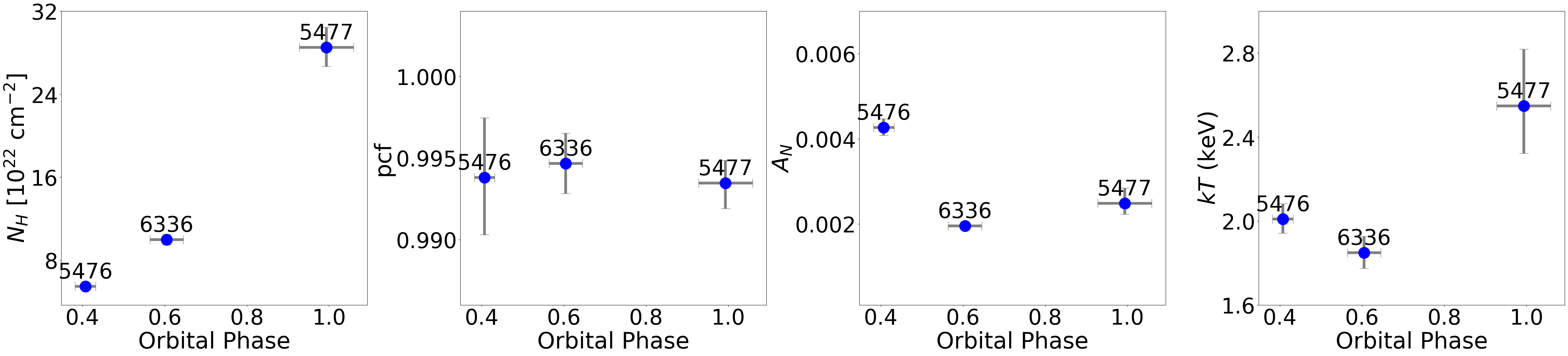}
    \caption{Variability of the best-fit continuum model parameters with orbital phase.}  
    \label{fig:params}
\end{figure*}

\paragraph{Orbital phase dependence of continuum parameters.}
The \texttt{Bmc} normalization, which reflects the intrinsic luminosity of the source and its distance \citep{1997ApJ...487..834T}, shows variations with orbital phase (see Fig.~\ref{fig:params}). This behavior may arise from intrinsic fluctuations in the accretion rate or from changes in the optical depth along the line of sight \citep{Levine2004}. \citet{Manikantan2023} reported a comparable case for the SgXB GX301$-$2, where orbital luminosity variations arise from changes in the mass accretion rate. Furthermore, variations in the orientation of the accretion column with respect to the observer can modulate the observed emission, as \citet{SanjurjoFerrn2020} showed for Cen~X$-$3.

We do not find a clear trend between the \texttt{Bmc} temperature ($kT$) and the orbital phase, although $kT$ reaches a peak in observation 5477, close to orbital phase 1.

\paragraph{Fe~K$\alpha$ evolution.}
The Fe~K$\alpha$ fluorescence line originates when hard X-ray photons (E > 7.1 keV) from the compact object ionize neutral or mildly ionized iron (Fe~\textsc{I–XVII}), ejecting a K-shell electron and leading to radiative recombination. This emission is generated in cold and dense regions of the stellar wind. The efficiency of this process depends on the column density $N_{\rm H}$ of neutral material along the line of sight, with higher $N_{\rm H}$ values leading to stronger fluorescence and larger EW \citep{George1991}. This trend has been observed in both persistent and transient HMXBs. Furthermore, \citep{https://doi.org/10.48550/arxiv.1501.03636} and \cite{Torrejn2010} showed that dense reprocessing material around the compact object enhances fluorescence, supporting the idea of a spherically distributed reprocessing region. However, in Fig. \ref{fig:EW_NH}, we observe a negative trend between the EW of the Fe~K$\alpha$ line and $N_{\rm H}$, contrary to our expectations.

\begin{figure*}[t]
    \centering
    \includegraphics[width=1\textwidth]{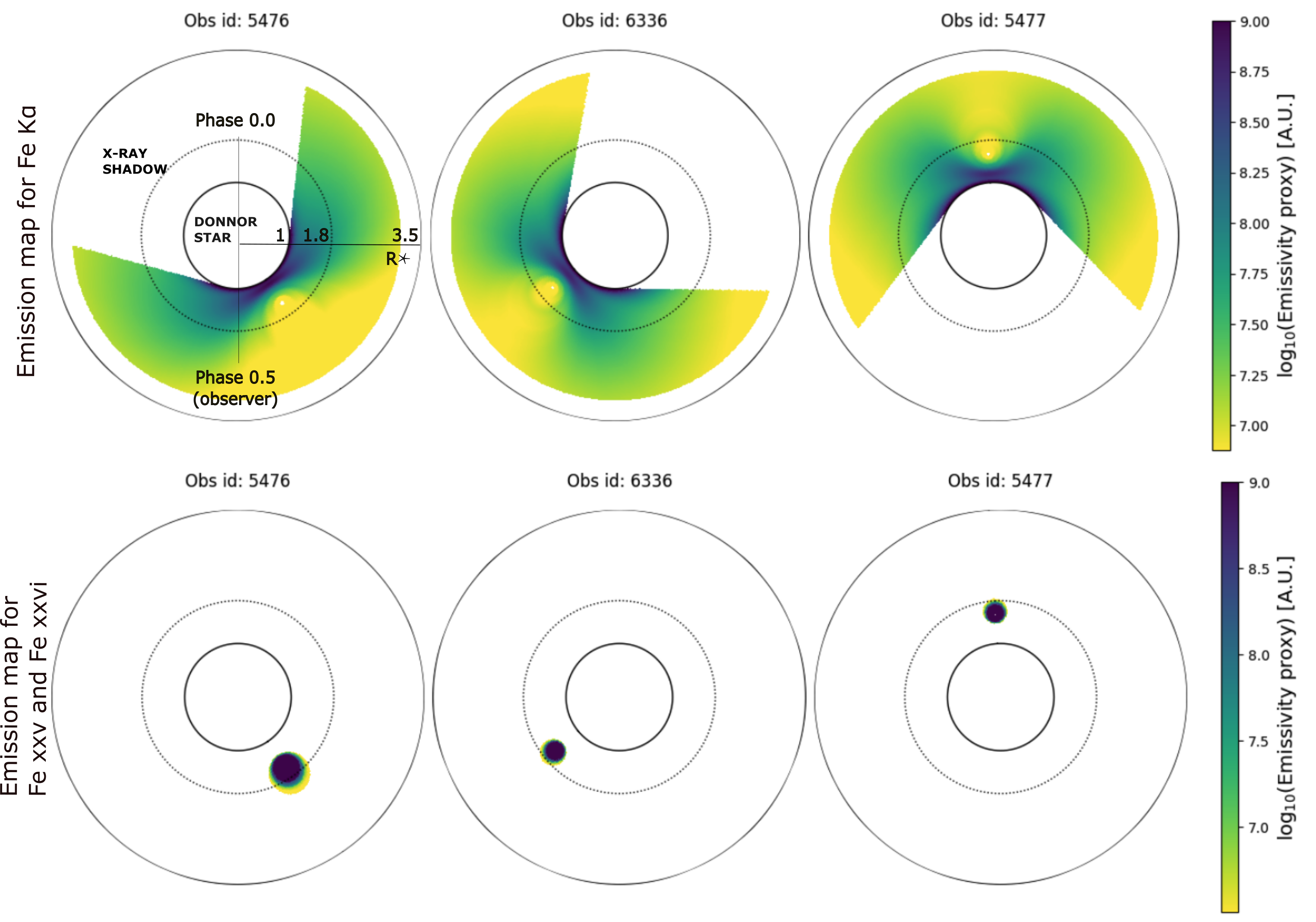}
    \caption{Upper row: Plane of the orbit of the relative emission region for Fe K$\alpha$. Bottom row:  Plane of the orbit of the relative emission region for the highly ionized Fe lines Fe \textsc{XXV} and Fe \textsc{XXVI} combined. The donor star radius and the orbital path are shown to scale. The color gradient represents the relative emission strength.} 
  \label{fig:ionization_maps}
\end{figure*}

This inverse relationship can be explained if the fluorescent line is formed mainly close to the surface of the donor star, where the wind density is high, particularly on the side illuminated by the NS. This interpretation is further supported by the ionization parameter map shown in Fig.~\ref{fig:ionization_maps}, where the distribution of $\log\xi$ highlights that the Fe~K$\alpha$ fluorescence site is located in the illuminated face of the donor star. When $N_{\rm H}$ is high, the emitting region becomes partially eclipsed from the observer’s viewpoint, while the continuum emitted by the NS remains largely visible, as absorption at 6.4 keV is relatively small. Consequently, the observed decrease in the Fe~K$\alpha$ line EW is primarily due to this geometric eclipse effect rather than absorption by the surrounding material.

Furthermore, the Fe~K$\alpha$ EW can vary with orbital phase due to changes in absorption and scattering conditions. When the compact object passes behind the donor, the EW may increase due to additional reprocessing by the wind \citep{Watanabe2006}. However, our results show a negative trend between the EW of the Fe~K$\alpha$ line and orbital phase. This can be explained using the same reasoning as above: because of the system’s inclination, the Fe~K$\alpha$ emitting region becomes increasingly obscured at orbital phases around 0 or 1, reducing the observed line strength relative to the continuum.

On the other hand, in Fig. \ref{fig:Fe_kalpha_vs_flux} right panel, is shown that the continuum flux exhibits a positive trend with the Fe~K$\alpha$ line flux, consistent with previous studies of other HMXBs by \citep{https://doi.org/10.48550/arxiv.1501.03636, Torrejn2010}.

\paragraph{Broadening of the Fe~K$\alpha$ emission line.}
When computing the FWHM velocity, we obtained values of $v_{\mathrm{FWHM}}\sim 2000-3000$ km s$^{-1}$ for observations 5476 and 6336 (Table~\ref{tab:line_kalpha}). These velocities are about one order of magnitude larger than both the $v_{\mathrm{orb}}$ and $v_{\infty}$, see Table~\ref{tab:stellar_wind_orbital_parameters}. 

In contrast, for observation 5477 the Fe line width is unresolved and, thus,  $v_{\mathrm{FWHM}} = 700$ km s$^{-1}$, which, although slightly larger, is of the same order of magnitude as $v_{\mathrm{orb}}$ and $v_{\infty}$.

These results have implications for the hypothesis that the Fe~K$\alpha$ line is formed close to the donor star. If that were the case, its width should be consistent with the local wind velocities. However, the observed discrepancy —where the line width exceeds the velocities expected from the stellar wind— indicates the action of additional broadening mechanisms. This difference can be explained if the Fe K$\alpha$ emission does not originate from a localized region, but instead forms throughout the extended stellar wind, an environment where velocities vary widely. Consequently, the profile we observe is the result of the superposition of contributions from all emitting regions, each Doppler-shifted by a different wind velocity. This explains both the broadened line widths and why the line centroid is not a reliable tracer of the system's orbital motion. Beyond this integrated wind broadening, several more specific processes have been proposed and observed in HMXBs that can further broaden or distort the line profile: (i) Compton scattering, which produces a red Compton shoulder and contributes to line broadening \citep{Watanabe2003}; this feature is also detected as tentative in the observations 5476; (ii) line blending and Doppler shifts from material with velocities of order $\sim 1000$ km/s \citep{gimenez2015}; and (iii) structured or turbulent stellar winds and outflows, which can produce asymmetric wings in the profile \citep{Torrejn2010}. Turbulent motions within the accretion stream might locally increase wind velocities to values capable of producing the observed broadening, although the physical properties of such flows are not yet well constrained. An alternative explanation is that the Fe K$\alpha$ emission does not arise from a single transition, but rather from the superposition of multiple K$\alpha$ lines associated with iron at different ionization stages \citep{Kallman2004}. In environments with lower $N_{\rm H}$ (and therefore less absorption), the relative contribution of all these broadening processes may become dominant, justifying line widths larger than those expected from the stellar wind velocities alone.

\paragraph{Compton Shoulder.}
The tentative presence of a Fe~K$\alpha$ CS in observation 5476 may indicate a dense stellar wind. The observed energy difference between the Fe~K$\alpha$ line and the CS is smaller than the maximum energy shift due to Compton back-scattering ($\theta=180^\circ$) for 6.4 keV photons:
$\Delta E_{\rm max} = 2 E_0^2 / (m_e c^2 + 2 E_0) \approx 0.156~\rm keV$.
Since the observed shoulder centroid is at 6.29 keV, $\Delta E = 0.11~\rm keV < \Delta E_{\rm max}$.

This supports the interpretation that the shoulder is produced through single Compton scattering of Fe~K$\alpha$ photons. Its shape and width are governed by the number of scattering events undergone by photons, becoming more pronounced in high-opacity environments \citep{Matt2002}. However, in our case, the observations with higher $N_{\rm H}$ do not reveal any feature even at a tentative significance level. 

This apparent discrepancy may be related to geometrical effects or to the physical conditions of the scattering medium. \citet{Odaka2016} studied the sensitivity of the Fe~K$\alpha$ CS to the geometry and variability of the X-ray illumination. Their simulations showed that the CS profile strongly depends on the inclination angle of the reflecting slab, except when the slab is Compton-thick. This implies that the observed EW of the CS can vary with the viewing angle, potentially producing smaller EWs in certain orientations despite a higher $N_{\rm H}$. Furthermore, the distribution and state of the scattering electrons, whether free or bound, can influence the CS profile. Factors such as electron thermal motion and binding effects in cold matter can blur the profile, similarly to thermal broadening in a plasma. Therefore, the observed discrepancy could result from a combination of geometric effects, viewing angle, and the physical state of the scattering electrons.

\paragraph{Fe~\textsc{XXV} and Fe~\textsc{XXVI}}.
We searched for the presence of highly ionized iron lines, in particular the Fe~\textsc{XXV} He-like triplet and the Fe~\textsc{XXVI} H-like line, in the spectra of the different observations. However, none of these features are detected with statistical significance. This is supported by the fact that Fe~\textsc{XXV} and Fe~\textsc{XXVI} ions are generated in much smaller volumes in the vicinity of the NS while the Fe k$\alpha$ feature is potentially emitted from a significant larger region in the circumstellar wind (see Fig.  \ref{fig:ionization_maps}).

\paragraph{Ionization state of the stellar wind.}
Our spectral analysis shows that the accretion wind in X1908+075 is characterized by neutral or low-ionization components, which are traced by the fluorescent Fe K$\alpha$ (6.4 keV) and K$\beta$ (7.1 keV) lines. Although highly ionized Fe lines such as Fe \textsc{XXV} and Fe \textsc{XXVI} are not detected in our spectra, their formation is expected under the high-ionization conditions present close to the NS \citep{Ebisawa1996}. 

We present emission line proxy maps of Fe species in the orbital plane (Fig.~\ref{fig:ionization_maps}). Adopting the CAK prescription for a smooth stellar wind and the X-ray luminosity of the NS, it is possible to calculate the density and ionization ($\xi$) at each point:
\begin{align}
v(r) &= v_\infty \left(1 - \frac{R_\star}{r}\right)^\beta, &
\rho(r) &= \frac{\dot{M}}{4\pi r^2 v(r)}, &
\xi(\mathbf{r}) &= \frac{L_{\rm X}}{n_e(\mathbf{r})\, r_{\rm X}^2},
\end{align}
where $v_\infty$ is the terminal velocity, $R_\star$ the stellar radius, $\beta$ the wind-acceleration parameter, $\dot{M}$ the mass-loss rate, $n_e$ the electron number density, $L_{\rm X}$ the X-ray luminosity of the compact object, with values taken from Table~\ref{tab:continuum}, and $r_{\rm X}$ the distance from the X-ray source to the plasma element. We used \texttt{xstar}\footnote{\url{https://heasarc.gsfc.nasa.gov/docs/software/xstar/docs/sphinx/xstardoc/docs/build/html/index.html}} \citep{2021Atoms...9...12M} to obtain the ionic fraction $f_{\rm ion}(\xi)$ as a function of $\log\xi$ in the thin-plasma approximation. The resulting map, computed for each point as
\begin{equation}
\label{eq:emissivity_proxy}
E_{\rm rel}(\mathbf{r}) \propto n_e(\mathbf{r})\, f_{\rm ion}\!\left(\xi(\mathbf{r})\right),
\end{equation}
serves as an emissivity proxy to trace the relative spatial distribution of line formation as a function of the system configuration.

\begin{figure}[t]
    \centering
    \includegraphics[width=1\columnwidth]{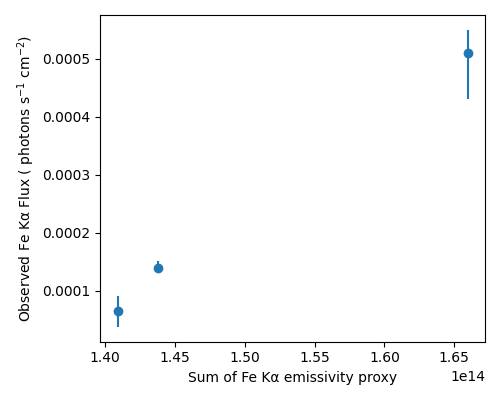}
    \caption{Integrated Fe K$\alpha$ emissivity proxy as shown in Fig. \ref{fig:ionization_maps} vs the observed X-ray emission line flux.} 
  \label{fig:feka_em_flux}
\end{figure}

The maps show that the Fe K$\alpha$ emission arises throughout the stellar wind, in an extended region between the NS and the donor star, while highly ionized Fe lines would form primarily close to the NS (with $\log \xi \sim 3$--4), extending out to roughly 1~$R_\star$. This picture is reinforced by the positive trend between the integrated Fe K$\alpha$ emissivity proxy for each observation, as shown in Fig.~\ref{fig:ionization_maps}, and the line flux derived from the X-ray spectra (see Fig.~\ref{fig:feka_em_flux}).

\paragraph{Fe~K$\beta$/K$\alpha$ line ratio and its flux dependence.}  
The theoretical Fe~K$\beta$/K$\alpha$ ratio lies in the range 0.11–0.17 \citep[see Fig.~2 in][]{palmeri2003modeling}, although observational measurements can deviate due to self-absorption and line-of-sight effects. In observation 6336, we found a higher ratio of $0.6\pm0.4$. The presence of a K-edge from near-neutral iron at $1.742\,\text{\AA}$ can alter the continuum near the Fe~K$\beta$ line \citep{vanderMeer2005}, potentially influencing the measured line strength. Similar effects have been observed in sources such as Vela X-1 \citep{Frst2013}. Although this K-edge is not detected in our spectral model—likely due to the limited resolution of {\it Chandra} HETGS—its presence could artificially enhance the apparent Fe~K$\beta$ flux, leading to an overestimation of the K$\beta$/K$\alpha$ ratio.

\section{Conclusions}

We have presented a spectral study of the SgXB X1908+075 based on three \textit{Chandra} HETGS observations obtained at different orbital phases. Our analysis of the iron K band provides new constraints on the structure, ionization state, and geometry of the stellar wind and its interaction with the NS. Our main conclusions are as follows:

\begin{enumerate}
\item Continuum variability and system parameters. The X-ray continuum is well described by a partially covered Comptonization model (\texttt{Bmc}). The hydrogen column density shows orbital modulation, peaking at $N_{\mathrm{H}} \approx 28.5 \times 10^{22}$ cm$^{-2}$ near $\phi_{\mathrm{orb}} \approx 1$. The intrinsic luminosity varies by a factor of $\sim 2$ between observations, reflecting changes in the mass accretion rate. Modeling this orbital modulation using a smooth CAK wind yields an inclination $i = 46 \pm 3^\circ$ and a donor mass-loss rate $\dot{M}_W = (9.1 \pm 1.6) \times 10^{-7}  M_\odot  \mathrm{yr}^{-1}$, consistent within a factor of $\sim 3$ with independent infrared estimates.

\item Statistical significance of emission lines. After accounting for the look-elsewhere effect through Monte Carlo simulations, only the Fe K$\alpha$ fluorescence line at 6.4 keV is robustly detected ($\sim3\sigma$ global significance) in all observations. Tentative features include a Compton shoulder in the least absorbed observation and Fe K$\beta$ in one observation.

\item Compton shoulder. A Compton shoulder associated with the Fe K$\alpha$ line is tentatively detected in the least absorbed observation, with an energy shift ($\Delta E \approx 0.11$ keV) consistent with single Compton scattering. This points to the presence of dense reprocessing material in the system. Its absence in higher-$N_{\mathrm{H}}$ observations may reflect geometrical effects, changes in the ionization state of the scattering medium, or dependence on viewing angle.

\item The EW of the Fe K$\alpha$ line anticorrelates with both column density and orbital phase, deviating from the canonical ``curve of growth''. This behavior can be explained if a significant fraction of the fluorescent emission originates near the illuminated face of the donor star and becomes partially occulted at phases of high absorption, while the continuum remains largely unaffected at 6.4 keV.

\item The Fe K$\alpha$ line is significantly broadened beyond what is expected from orbital or terminal wind velocities. The measured FWHM velocities ($>10^3$ km s$^{-1}$) suggest that the line forms over an extended region of the stellar wind with a wide velocity distribution, rather than in a single localized site. Additional contributions from Compton scattering, turbulence, and line blending are likely present.

\end{enumerate}
Future high-sensitivity observations covering a wider range of orbital phases, for example with \textit{XRISM} and \textit{Athena}, will allow a more precise separation of the effects of fluorescence, scattering, and ionization on the iron line emission in this source.

\section*{Data availability}
The data analyzed in this study can be found in the {\it Chandra} archive under the observation identification numbers 5476, 6336, and 5477 (\url{https://cda.harvard.edu/chaser/searchOcat.action}).
\begin{acknowledgements}

The author(s) acknowledge the financial support from the MCIN with funding from the European Union NextGenerationEU and Generalitat Valenciana in the call Programa de Planes Complementarios de I+D+i (PRTR 2022). Project (Athena-XIFU-UA), reference ASFAE/2022/002. We would like to thank the anonymous referee for their helpful criticism. 
J.J.R.R. acknowledges financial support from the Spanish Ministry of Education, Culture and Sport fellowship PRX23/00270, and also thanks all the staff from SRON for their collaboration and hospitality there. Part of this work has been developed in the framework of the PID2024-155779OB-C33 project funding by MICIU/AEI/10.13039/501100011033 and by FEDER (UE).
\end{acknowledgements}

\bibliographystyle{aa}
\bibliography{example.bib} 
\newpage

\begin{appendix}
\onecolumn

\section{Soft-excess modeling with \texttt{Blackbody}}
\label{appendix:soft_excess}

We summarize here the different spectral configurations tested to model the soft excess in the 0.5–2 keV range using an additional \texttt{Blackbody} component.

We explored three absorption scenarios for the \texttt{Blackbody}: (i) an independent absorption column $N_{\rm H}(BB)$ constrained to the interstellar range (1.17–1.26) $\times 10^{22}$ cm$^{-2}$ estimated with the HEASARC tool\footnote{\url{https://heasarc.gsfc.nasa.gov/cgi-bin/Tools/w3nh/w3nh.pl}}, (ii) a freely varying $N_{\rm H}(BB)$, and (iii) a configuration where the \texttt{Blackbody} shares the same absorption column as the \texttt{Bmc} component.

The resulting best-fit parameters for each configuration are reported in Table~\ref{tab:soft_excess}, including $N_{\rm H}$, $kT$, normalization values for both components, the unabsorbed \texttt{Blackbody} flux, and the flux ratio relative to the \texttt{Bmc} component. The same set of configurations was also tested replacing the \texttt{Blackbody} with alternative thermal models (\texttt{mekal} and \texttt{diskbb}), yielding consistent results.

\begin{table*}[h]
\centering
\caption{Summary of the \texttt{Blackbody} soft-excess modeling for three configurations.}
\label{tab:soft_excess}
\begin{adjustbox}{max width=\textwidth}
\begin{tabular}{l l c c c c c c c c c}
\hline
Obs. ID & Configuration & $N_{\rm H}(\texttt{Bmc})$ & $A_{N}$ & $kT_{\texttt{Bmc}} $ (keV) & $N_{\rm H}(\texttt{BB})$ & Norm (\texttt{BB}) & $kT_{(\texttt{BB})}$ (keV)& $F_{\rm unabs}$ & Flux ratio (\texttt{BB}/\texttt{Bmc}) & $\chi^2_{\rm red}$ \\
\hline

5476 & $N_{\rm H}$ ISM & $6.7^{+0.7}_{-0.4}$ & $<0.18$ & $1.5^{+0.4}_{-0.3}$ & $1.26^{+0.00}_{-0.10}$ & $\left(4^{+2}_{-4}\right)\times10^{-5}$ & $1.0^{+0.0}_{-0.8}$ & $3.49\times10^{-12}$ & $1.18\times10^{-2}$ & 1.21 \\

5476 & $N_{\rm H}$ free & $4.2^{+0.9}_{-0.5}$ & $<0.4$ & $2.7\pm0.7$ & $10^{+18}_{-2}$ & $\left(7\pm4\right)\times10^{-4}$ & $0.94^{+0.06}_{-0.00}$ & $6.15\times10^{-11}$ & $2.57\times10^{-1}$ & 1.20 \\

5476 & $N_{\rm H}$ same as \texttt{Bmc} & $5.4^{+0.5}_{-0.4}$ & $\left(5^{+8}_{-1}\right)\times10^{-3}$ & $1.9^{+0.2}_{-0.3}$ & - & $<10$ & $<1$ & $0$ & $0$ & 1.22 \\

\hline

6336 & $N_{\rm H}$ ISM & $9.3\pm0.5$ & $<40$ & $1.90^{+0.08}_{-0.09}$ & $1.17^{+0.10}_{-0.00}$ & $0.010^{+0.013}_{-0.007}$ & $<2.2$ & $7.96\times10^{-12}$ & $6.12\times10^{-2}$ & 1.19 \\

6336 & $N_{\rm H}$ free & $9.3\pm0.5$ & $\left(1.98^{+0.03}_{-0.08}\right)\times10^{-3}$ & $1.90^{+0.08}_{-0.09}$ & $\le0.14$ &  $\left(7.4\pm2.4\right)\times10^{-6}$ & $0.101^{+0.015}_{-0.008}$ & $1.06\times10^{-13}$ & $8.12\times10^{-4}$ & 1.19 \\

6336 & $N_{\rm H}$ same as \texttt{Bmc} & $9.3^{+0.4}_{-0.3}$ & $<10$ & $1.90^{+0.08}_{-0.09}$ & - & $<10$ & $0.001$ & $0$ & $0$ & 1.23 \\

\hline

5477 & $N_{\rm H}$ ISM & $28.9^{+2.0}_{-1.9}$ & $\left(2.5\pm 0.3\right)\times10^{-3}$ & $2.5^{+0.3}_{-0.2}$ & $1.26^{+0.00}_{-0.10}$ & $\left(3.7^{+0.7}_{-1.3}\right)\times10^{-6}$ & $1.00^{+0.00}_{-0.19}$ & $2.95\times10^{-13}$ & $2.43\times10^{-3}$ & 1.19 \\

5477 & $N_{\rm H}$ free & $31\pm 0.3$ & $\left(2.4^{+0.3}_{-0.2}\right)\times10^{-3}$ & $2.4^{+0.3}_{-0.2}$ & $3.6^{+2.0}_{-1.4}$ & $\left(10^{+8}_{-6}\right)\times10^{-6}$ & $1.0^{+0.0}_{-0.5}$ & $8.39\times10^{-13}$ & $6.69\times10^{-3}$ & 1.15 \\

5477 & $N_{\rm H}$ same as \texttt{Bmc} & $26.5^{+2.2}_{-1.5}$ & $\left(2.6^{+0.5}_{-0.3}\right)\times10^{-3}$ & $2.7\pm 0.3$ & - & $10^{+0}_{-8}$ & $0.113^{+0.015}_{-0.005}$ & $2.53\times10^{-8}$ & $2.13\times10^{2}$ & 1.27 \\

\bottomrule
\end{tabular}
\end{adjustbox}
\tablefoot{Fluxes are in erg\,s$^{-1}$\,cm$^{-2}$.}
\end{table*}

\section{Blind line search}
\label{appendix:linesearch}

Here we provide a more detailed description of the blind line search procedure and the Monte Carlo assessment of statistical significance.

We performed a blind search for narrow spectral features using a grid-search technique in wavelength space, following standard high-resolution X-ray spectroscopy approaches (e.g., \citealt{Pinto2017, Kosec_2018}). Starting from the best-fit continuum model, we introduced a narrow Gaussian component with fixed centroid and width, allowing only the normalization to vary (positive and negative values to account for emission and absorption).

\begin{figure*}[h!] 
    \centering 

    \includegraphics[width=0.99\textwidth]{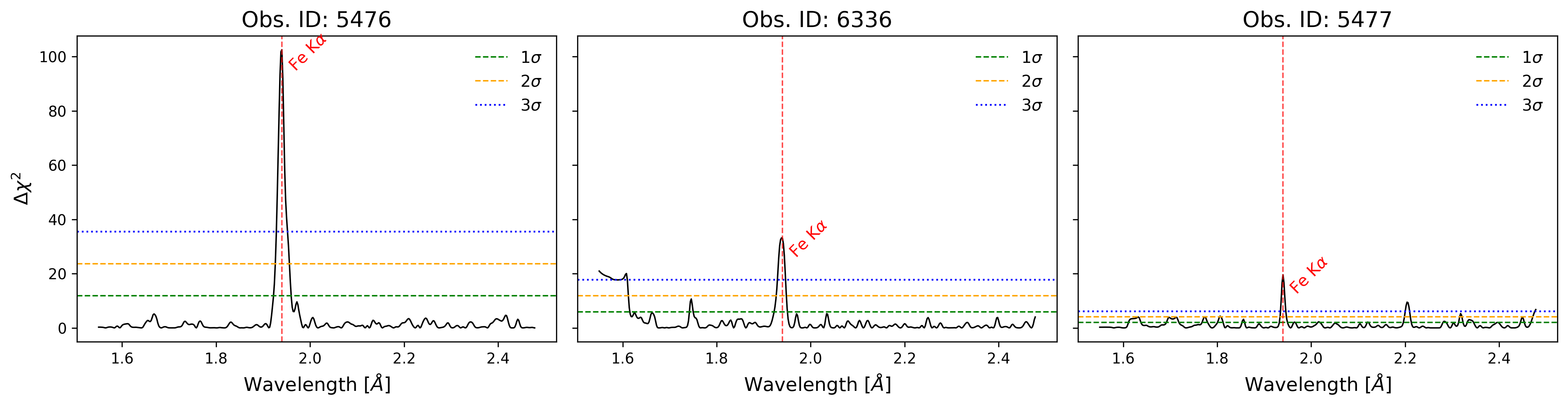} 
    \caption{$\Delta \chi^2$ as a function of wavelength obtained from the blind line search. The maximum $\Delta \chi^2$, highlighted in red, corresponds to the Fe K$\alpha$ emission line and exceeds the $3\sigma$ significance level. Additional candidate features are visible at the $1\sigma$ and $2\sigma$ levels. Note that the significances shown here are before accounting for the look-elsewhere effect; therefore, they differ from the final values reported.}
    \label{fig:linesearch} 
\end{figure*}

\begin{figure*}[h!] 
    \centering 
    \includegraphics[width=\textwidth]{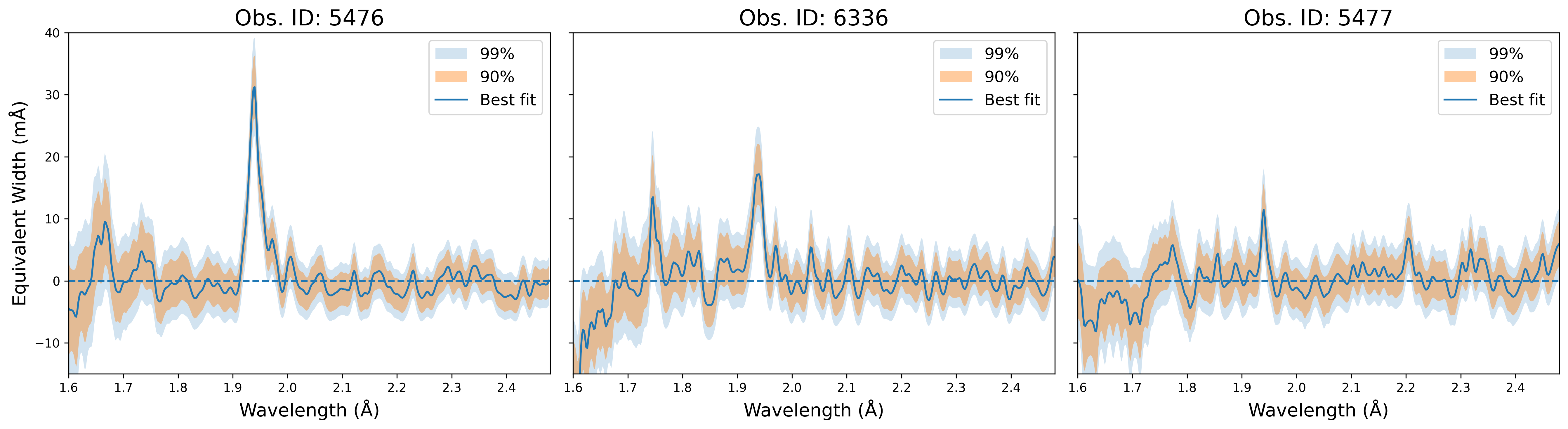} 
    \caption{EW scan as a function of wavelength. The dark blue curve shows the best-fit EW obtained by stepping a narrow Gaussian across the spectrum. The shaded regions represent the 90\% (orange) and 99\% (light blue) confidence intervals derived from the Gaussian line flux uncertainties. The horizontal dashed line marks EW = 0 m$\AA$.}
    \label{fig:ew_ma}
\end{figure*}

The centroid was stepped across the 1.55–2.48 $\AA$ wavelength range (corresponding to the Fe K band, 5–8 keV) in increments of 0.002 $\AA$, comparable to the instrumental highest spectral resolution in this band ($\sim$ 0.002–0.005 $\AA$)\footnote{\url{https://cxc.harvard.edu/cal/}}. At each step, we computed the improvement in fit statistic, $\Delta \chi^2$, with respect to the continuum-only model and measured the corresponding EW. The resulting $\Delta \chi^2$ and EW profiles are shown in Figs.~\ref{fig:linesearch} and \ref{fig:ew_ma}.

To assess the global significance of candidate features and account for the look-elsewhere effect, we performed Monte Carlo simulations by generating 1000 synthetic spectra with the \texttt{fakeit} command. The simulations were based on the best-fit continuum model and included the same instrumental responses, exposure times, and background as the real data.

For each simulated spectrum, we applied the same blind grid-search procedure as used for the data and recorded the maximum improvement in the fit statistic, $\Delta \chi_{\rm max,sim}^{2}$.

\begin{figure*}[h!]
\centering
\includegraphics[width=\textwidth]{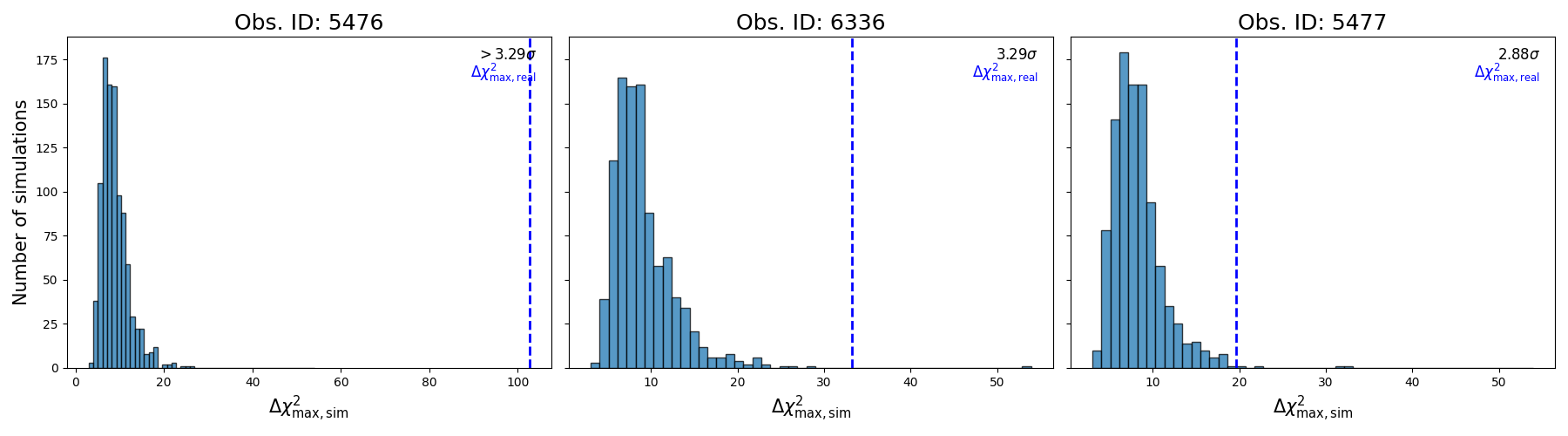}
\caption{Distribution of the maximum $\Delta \chi^2$ values obtained from 1000 Monte Carlo simulations under the null hypothesis (no spectral lines) for the three observations. The blue dashed vertical line in each panel marks the observed $\Delta \chi^2_{\rm max,real}$ measured in the real data. The fraction of simulations with $\Delta \chi_{\rm max,sim}^{2} \ge \Delta \chi^2_{\rm real}$ defines the global $p$-value, accounting for the look-elsewhere effect. The corresponding global significance (in Gaussian $\sigma$ units) is indicated in each panel.}
\label{fig:mc_dist}
\end{figure*}

The significance of observed features was evaluated by comparing each $\Delta \chi_{\rm real}^{2}$ with the distribution of simulated maxima. The corresponding global $p$-value was computed as
\begin{equation}
p_i = \frac{N(\Delta \chi_{\rm max,sim}^{2} \ge \Delta \chi_{i~\rm real }^{2})}{N_{\rm sim}},
\end{equation}
where $N_{\rm sim} = 1000$ and $i$ labels each feature. The resulting distribution of $\Delta \chi_{\rm max,sim}^{2}$ and the observed values are shown in Fig.~\ref{fig:mc_dist}.

Global $p$-values were converted to Gaussian significances using
\begin{equation}
\sigma = \sqrt{2}\,\mathrm{erf}^{-1}(1 - 2p),
\end{equation}
where $\mathrm{erf}^{-1}$ is the inverse error function.

A detection threshold of $p < 0.005$ was adopted to define statistically significant features. Gaussian modeling of the lines and their best-fit parameters are reported in Table~\ref{tab:lines}.

\newpage
\twocolumn

\section{Description of the \texttt{fit\_nh\_ps} column-density fitting routine}
\label{appendix:nh_ps_fit}
The hydrogen column density, $N_{\rm H}$, encountered by the emerging X-ray radiation depends on the density structure of the donor-star wind and on the relative geometry of the compact object, the donor star, and the observer. Consequently, the observed $N_{\rm H}$ provides valuable constraints on the wind properties and the orbital configuration of the system.

The hydrogen column density along the line of sight is computed by integrating the wind density between the compact object and the observer,
\begin{equation}
N_{\rm H} = \int_{l_0}^{\infty} n_{\rm H}(l)\, dl ,
\end{equation}
where $n_{\rm H}$ denotes the hydrogen number density. As a first-order approximation, we assume that $n_{\rm H}$ scales linearly with the wind density. The variable $l$ represents the coordinate along the line of sight. Under the assumption of a stationary, spherically symmetric stellar wind, the radial density profile can be expressed as
\begin{align}
\rho(r) &= \frac{\dot{M}}{4\pi r^2 v(r)},\\
v(r) &= v_\infty \left(1 - \frac{R_\star}{r}\right)^{\beta},
\end{align}
where $\dot{M}$ is the stellar mass-loss rate and $v(r)$ is the wind velocity at a distance $r$ from the donor star. The wind velocity is described using a standard $\beta$-law, where $v_\infty$ is the terminal wind velocity and $\beta$ determines the acceleration profile of the wind.

To constrain the system parameters, we fit the measured $N_{\rm H}$ values using the \texttt{fit\_nh\_ps} routine from the \texttt{xraybinaryorbit}  package \citep{SanjurjoFerrn2024}, which employs particle swarm optimization (PSO) to minimize the difference between the observed and model-predicted column densities. PSO is a gradient-free global optimization method in which a population of candidate solutions (particles) explores the parameter space. This approach is well suited to non-linear problems with correlated parameters, where gradient-based methods may converge slowly or become trapped in local minima.

The orbital configuration is described by the semi-major axis ($a$), orbital period ($P_{\rm orb}$), eccentricity ($e$), argument of periastron ($\omega$), and inclination ($i$). The wind model is parameterized by the mass-loss rate ($\dot{M}$), terminal wind velocity ($v_\infty$), and the CAK velocity-law exponent ($\beta$). In our calculations, the semi-major axis is expressed in units of the stellar radius $R_\star$. To reduce parameter degeneracies, $R_\star$ was fixed to its most probable value, while the semi-major axis was allowed to vary within its plausible range. The remaining parameters were allowed to vary within physically motivated bounds taken from previous studies (Table~\ref{tab:stellar_wind_orbital_parameters}). The best-fit model reproduces the observed column densities reasonably well. For the three orbital phases considered, the predicted values of $N_{\rm H}$ are $5.5$, $7.9$, and $27.3$ (in units of $10^{22}\mathrm{cm^{-2}}$), compared to the observed values of $4.9$, $8.8$, and $27.3$, respectively (the complete $N_{\rm H}$ column minus the interstellar absorption with respect to the source). The corresponding best-fit orbital and wind parameters are listed in Table~\ref{tab:orbital_params_ps}. 

\begin{figure}[ht]
    \centering
    \includegraphics[width=0.99\columnwidth]{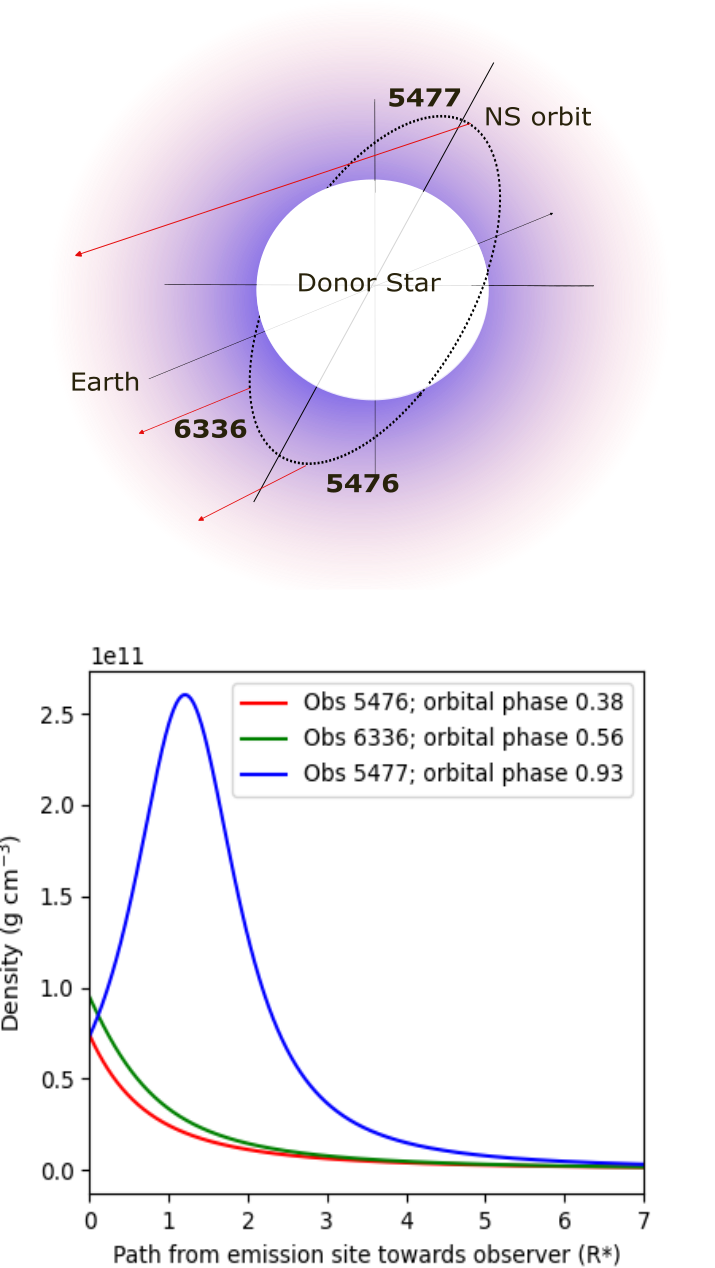}
    \caption{Upper panel: schematic representation of the binary system and the corresponding lines of sight toward the observer for the three observations. Lower panel: density profiles traversed by the X-ray radiation emitted by the NS along the line of sight for each observation.}
    \label{fig:nh_appendix}
\end{figure}

The upper panel of Fig. ~\ref{fig:nh_appendix} shows a schematic representation of the binary system and the corresponding lines of sight toward the observer for the three observations. The three lower panels represent the density profile traversed by the x-ray radiation from the NS in its path towards the observer for each observation.
\begin{table}[h]
\caption{Best-fit orbital and stellar wind parameters obtained from the PSO optimization.}
\label{tab:orbital_params_ps}
\centering
\begin{tabular}{lc}
\hline\hline
Parameter & Value \\
\hline
Semi-major axis, $a$ ($R_\star$) & $1.70 \pm 0.06$ \\
Orbital period, $P_{\rm orb}$ (days) & $4.4008 \pm 0.0005$ \\
Eccentricity, $e$ & $0.060 \pm 0.018$ \\
Argument of periastron, $\omega$ (deg) & $269 \pm 11$ \\
Inclination, $i$ (deg) & $46 \pm 3$ \\
Mass-loss rate, $\dot{M}$ ($M_\odot\,\mathrm{yr}^{-1}$) & $(9.1 \pm 1.6)\times10^{-7}$ \\
Terminal wind velocity, $v_\infty$ (km s$^{-1}$) & $490 \pm 90$ \\
Velocity-law exponent, $\beta$ & $1.210 \pm 0.005$ \\
\hline
\end{tabular}
\end{table}

\end{appendix}

\end{document}